# Tunable Interband Transitions in Twisted *h*-BN/Graphene Heterostructures

Bingyao Liu[1,2,#], Yu-Tian Zhang,[3,#] Ruixi Qiao,[4,5,#] Ruochen Shi,[1,4] Yuehui Li,[1,4] Quanlin Guo,[6] Jiade Li,[3,7] Xiaomei Li,[1,4] Li Wang,[7] Jiajie Qi,[6] Shixuan Du,[3,8] Xinguo Ren,[3] Kaihui Liu,[4,6,8,9,*] Peng Gao,[1,2,4,9,10,*] and Yu-Yang Zhang.[3,*]

[1] *Electron Microscopy Laboratory, School of Physics, Peking University, Beijing 100871, China;*

[2] *Academy for Advanced Interdisciplinary Studies, Peking University, Beijing 100871, China;*

[3] *University of Chinese Academy of Sciences and Institute of Physics, Chinese Academy of Sciences, Beijing 100049, China;*

[4] *International Center for Quantum Materials, School of Physics, Peking University, Beijing 100871, China;*

[5] *Institute for Frontier Science, Nanjing University of Aeronautics and Astronautics, Nanjing 210016, China;*

[6] *State Key Laboratory for Mesoscopic Physics, Frontiers Science Center for Nano-optoelectronics, School of Physics, Peking University, Beijing 100871, China*

[7] *Beijing National Laboratory for Condensed Matter Physics, Institute of Physics, Chinese Academy of Sciences, Beijing 100190, China*

[8] *Songshan Lake Materials Laboratory, Dongguan, Guangdong 523808, China;*

[9] *Collaborative Innovation Center of Quantum Matter, Beijing 100871, China;*

[10] *Interdisciplinary Institute of Light-Element Quantum Materials and Research Center for Light-Element Advanced Materials, Peking University, Beijing 100871, China.*

[#] B.Y. Liu, Y.T. Zhang and R.X. Qiao contributed equally to this work.

[*] E-mail: khliu@pku.edu.cn; p-gao@pku.edu.cn; zhangyuyang@ucas.ac.cn.

**Abstract**

In twisted *h*-BN/graphene heterostructures, the complex electronic properties of the fast-traveling electron gas in graphene are usually considered to be fully revealed. However, the randomly twisted heterostructures may also have unexpected transition behaviors, which may influence the device performance. Here, we study the twist angle-dependent coupling effects of *h*-BN/graphene heterostructures using monochromatic electron energy loss spectroscopy. We find that the moiré potentials alter the band structure of graphene, resulting in a redshift of the intralayer transition at the M-point, which becomes more pronounced up to 0.25 eV with increasing twist angle. Furthermore, the twisting of the Brillouin zone of *h*-BN relative to the graphene M-point leads to tunable vertical transition energies in the range of 5.1-5.6 eV. Our findings indicate that twist-coupling effects of van der Waals heterostructures should be carefully considered in device fabrications, and the continuously tunable interband transitions through the twist angle can serve as a new degree of freedom to design optoelectrical devices.

**Introduction**

Two-dimensional (2D) materials have various electronic structures and can be artificially assembled, drawing great attention since the first exfoliation of graphene[1]. When stacked together, there will be no strict limitation in lattice matching during heterostructure assembling due to the van der Waals interaction in 2D materials[2], bringing possibilities for many complex structures[3–5]. The twist angle between these stacked 2D layers also supplies a new degree of freedom to perform bandgap engineering[6–8]. The tunable interfacial interactions offer platforms to study many interesting phenomena, e.g., the continuously tunable van Hove singularities (vHS)[9–13] and subsequent superconductivity[14] in twisted bilayer graphene, second-generation Dirac cones in hexagonal boron nitride ($h$-BN)/graphene heterostructures[15,16], and moiré excitons in twisted transition-metal dichalcogenides[17,18].

In van der Waals heterostructure optoelectronic devices, $h$-BN and graphene are often used as dielectric encapsulation layers[19] and electrodes[20,21], respectively. $h$-BN usually serves as a substrate or encapsulation layer to hold other fragile 2D materials or protect them from exposure to the environment and avoid oxidation or contamination due to its stability and dangling bond-free property[17,18,22,23], while it has a minor influence on the intrinsic performance of the devices due to its large bandgap[24]. As a 2D semimetal material, graphene has a flat surface and zero bandgap[25,26] and is thus suitable for electrical contact in 2D devices[27]. However, the contact of graphene and $h$-BN creates 2D heterointerfaces, in which the interfacial moiré potential between them may influence the band structure[28]. Especially at some specific twist angles, the overlapped bands of the stacked layers created new Van Hove singularities without symmetry protection, and some unexpected interlayer vertical transitions may also influence the electrical and optical properties of the devices.

Herein, by using scanning transmission electron microscopy-electron energy loss spectroscopy (STEM-EELS) and first-principle calculations, we find that the interband transition behavior in $h$-BN/graphene van der Waals heterostructures exhibits a strong twist angle dependence. The moiré potentials add interference to the energy bands of graphene, making the intralayer transition redshift as the twist angle increases. Meanwhile, since the reciprocal lattices of *h*-BN and graphene also twist concurrently as their lattices twist in real space, such twisting of the *h*-BN Brillouin zone (BZ) projected onto the graphene M-point leads to tunable vertical transition energies. Our study reveals the angle-dependent coupling behavior between *h*-BN and graphene, indicating that the coupling effects in 2D devices should be carefully considered to prevent unexpected transitions and/or transition energy shifts. Moreover, the revealed twist angle-related transition behavior in twisted *h*-BN/graphene enables tunable intralayer and interlayer transition energies, creating opportunities for the fabrication of new 2D optoelectrical devices[27,29] with artificially specified wavelengths.

**Results and discussion**

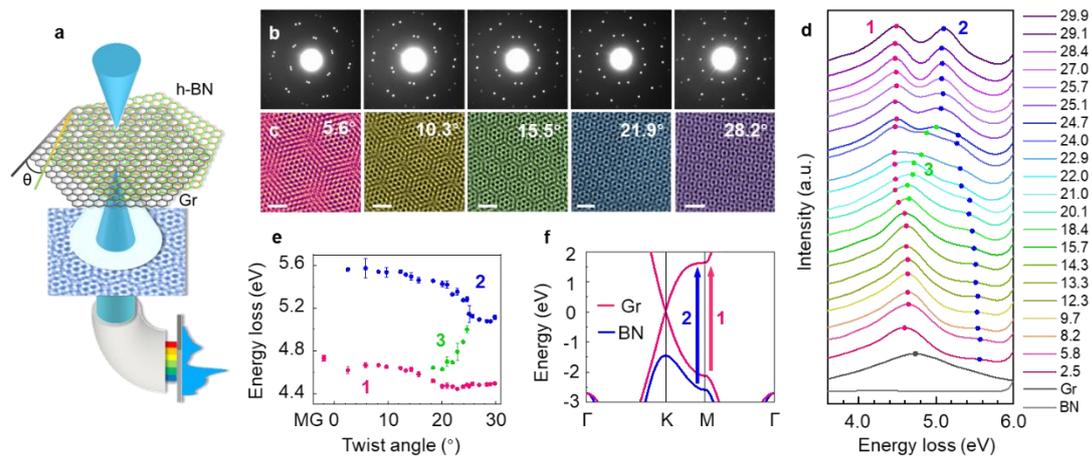

**Figure 1.** Twist angle-resolved STEM-EELS of *h*-BN/graphene (Gr) vertical heterostructures. (a) Schematic of the STEM-EELS measurement. (b-c) Diffraction patterns (b) and HAADF images (c) of *h*-BN/Gr with twist angles of 5.6°, 10.3°, 15.5°, 21.9° and 28.2°. Scale bars: 1 nm. (d) EELS spectra of *h*-BN/Gr with different twist angles. (e) Transition energies extracted from (d), in which MG corresponds to the monolayer graphene. (f) Calculated band structure of a *h*-BN/Gr heterostructure with a twist angle of 0°.

We used chemical vapor deposition (CVD)-synthesized graphene and *h*-BN

monolayers to assemble vertical heterostructures, which were subsequently transferred onto TEM grids for characterization. A schematic of the STEM-EELS measurement is shown in **Figure 1(a)**. High-angle annular dark field (HAADF) images and EELS spectra were recorded simultaneously. The former, together with the selected area electron diffraction (SAED) patterns, were employed to determine the twist angles. **Figure 1(b)** shows five typical SAED patterns with twist angles of 5.6°, 10.3°, 15.5°, 21.9° and 28.2°, where two sets of sixfold symmetry diffraction patterns corresponding to graphene and *h*-BN are rotated from each other and have slightly different lattice constants (the lattice of *h*-BN is ~1.8% larger than that of graphene[30]). The HAADF images taken in corresponding areas show different moiré periods [**Figure 1(c)**], which fit well with the results calculated from the twist angles extracted from SAED (**Figure S1**).

The typical EELS spectra of twist angle-dependent transitions in *h*-BN/graphene vertical heterostructures are shown in **Figure 1(d)**, whose peak positions are extracted and demonstrated in **Figure 1(e)**. The corresponding twist angles determined by SAED are shown in **Figure S2**. Three main peaks can be found in the range of 4-6 eV. Earlier studies suggested that the peak in graphene at approximately 5 eV can be recognized as π plasmons[31,32]. However, recent work noted that this excitation is actually an interband transition[33] in EELS characterizations. Especially in our experiment, the EELS collective semi-angle is compatible with the incident angle (see Methods in Supplemental Material for details), making the direct transmission electrons, i.e., q→0, prominent, while the contribution from the non-zero momentum transfer interaction is negligible (a detailed discussion can be found in the SM and **Figure S3**). Therefore, we will only consider vertical interband transitions hereinafter. As a comparison, we also performed EELS on monolayer graphene and *h*-BN [dark and light gray lines in **Figure 1(d)**, respectively]. Notably, the *h*-BN spectrum shows no peaks below 6 eV, and only a rising edge at 6 eV results from its bandgap, which also appears in the *h*-BN/graphene heterostructure spectra and remains unchanged regardless of the presence of graphene or the twist angle. The graphene spectrum has a single peak at ~4.75 eV, close to the peaks with the lowest energies in the twisted *h*-BN/graphene heterostructure spectra, and is labeled the 1st transition. Notably, the 2nd and 3rd transitions did not appear in the *h*-BN or graphene monolayer spectrum, but showed up in the *h*-BN/graphene heterostructure spectra, indicating that the new 2nd and 3rd transitions originate from the interlayer transitions between *h*-BN and graphene.

Combined with first-principle calculations, we identified the 1st and 2nd transitions as the graphene intralayer transition and the interlayer transition from *h*-BN to graphene [**Figure 1(f)**, the pink and blue arrows indicate the transition pathways of intralayer and interlayer transitions, respectively]. The hopping end points are both the M-point in the graphene conduction band, while the starting points can be attributed to the M-point in the valence bands of graphene and *h*-BN, respectively[34]. The 3rd transition might be caused by the overlapped bands from *h*-BN and graphene owing to the twist mechanism and will be discussed below.

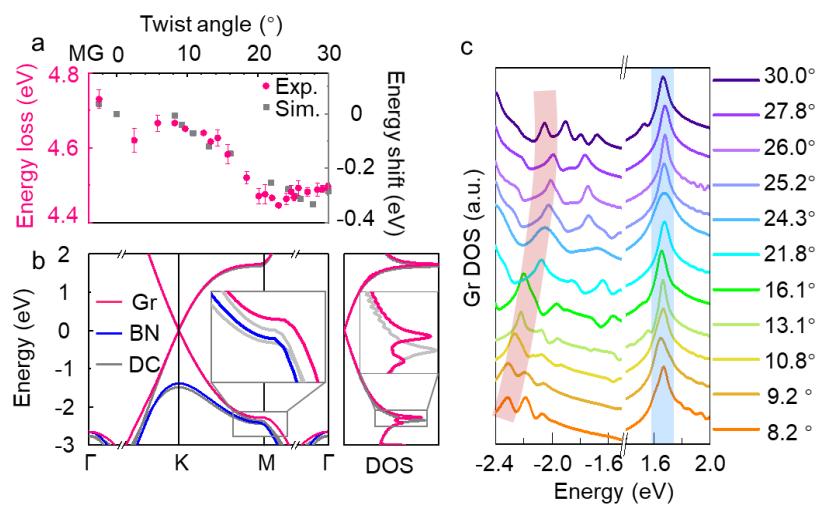

**Figure 2.** Graphene intralayer transition. (a) Comparison of experimental results (Exp.) and density functional theory (DFT)-simulated results (Sim.) regarding the Gr-Gr transition energy redshift with the presence of *h*-BN and with the twist angle increasing from 0° to 30°. (b) Calculated band structure and DOS of 0° twist coupled *h*-BN/Gr (blue and pink lines for *h*-BN and Gr, respectively) heterostructures and decoupled (DC, i.e., separated *h*-BN and graphene, gray) monolayers. (c) Calculated Gr DOS in *h*-BN/Gr heterostructures with different twist angles.

We first studied the evolution of the graphene intralayer transition at the M-point with the *h*-BN interaction. The statistics of all experimental results are shown as pink points in **Figure 2(a)**. The transition energy decreases by ~0.25 eV with the presence of *h*-BN and with the twist angle increasing from 0° to 30°. First-principle calculations were performed to reveal the underlying mechanism of the energy shift. A comparison of the band structure and density of state (DOS) of graphene between 0°-twisted coupled and decoupled *h*-BN/graphene heterostructures is shown in **Figure 2(b)**. We found two DOS peaks that rightly originate from the graphene vHS at the M-point in both the valence band and conduction band, contributing to the response in the EELS spectrum of the graphene intralayer transition. Interestingly, the interaction of graphene and *h*-BN valence bands is enhanced at the M-point due to the aggregation of electrons. The small gap at the M-point is enlarged by this enhanced interaction, pushing the graphene valence band up closer to the Fermi level and thus leading to a smaller intralayer hopping energy. Moreover, the interaction can cause interlayer wavefunction hybridization; thus, the graphene valence vHS at the M-point splits into two subpeaks [enlarged DOS in **Figure 2(b)**], corroborating the energy drop of the M-point intralayer hopping energy of graphene.

The DOS projected onto graphene in the *h*-BN/Gr heterostructures with different twist angles are plotted in **Figure 2(c)**. The pink shade marks the variation in the valence vHS of graphene, which is the starting point of intralayer hopping and decreases as the twist angle increases. At the same time, the conduction vHS (ending point) covered by the blue band remains unchanged with different twist angles. Therefore, the intralayer transition energy drop is caused by the variation in the valence band of graphene owing to a different moiré potential at the heterointerface. Note that although the well-known bandgap error with the Perdew-Burke-Ernzerhof (PBE) functional[35,36] makes the simulated transition energies lower than the experimental results, the tendency of the simulated energy shift [gray squares in **Figure 2(a)**] shows excellent agreement with the experimental results.

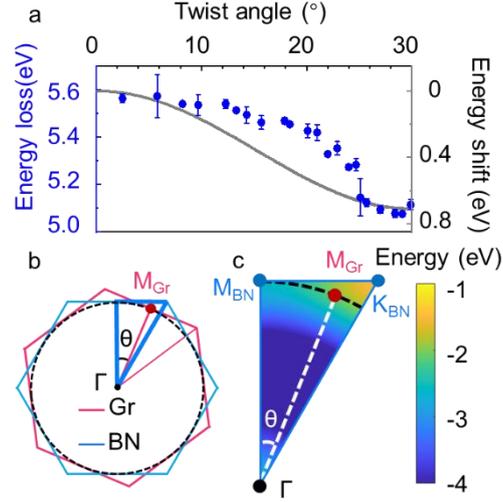

**Figure 3**. *h*-BN/Gr interlayer transition. (a) Experimental results (blue points) and DFT-simulated results (gray line) of the transitions with different twist angles. (b) Schematic of the θ-twist *h*-BN/Gr BZ, in which the black dashed circle indicates the track of the Gr M-point with all possible twist angles. (c) Extracted valence band energy of *h*-BN in the Γ-M-K triangle, in which the black dashed line indicates the path along which the graphene M-point projects in the *h*-BN BZ. Color bar: the relative energy of the *h*-BN valence band below the Fermi level.

The energy of the 2$^{nd}$ interlayer transition in the *h*-BN/graphene heterostructure was extracted and is represented as blue points in **Figure 3(a)**. The hopping energy decreases ~0.5 eV from 5.6 eV to 5.1 eV as the twist angle increases. Based on the above discussion, this transition occurs between the valence band of *h*-BN, which varies with the twist angle, and the flat M-point conduction band of graphene. Since the graphene M-point conduction band remains fixed, we only considered the evolution of the transition starting point on the *h*-BN valence band. The BZ of the twisted system is schematically shown in **Figure 3(b)**, where the pink and blue hexagons correspond to the BZ for graphene and *h*-BN, respectively. When the twist angle θ changes, the graphene M-point will project onto different positions in the *h*-BN BZ. In this way, the starting points in the interlayer vertical transitions will form a circular track in the *h*-BN BZ with a radius of the Γ-M reciprocal distance when θ varies from 0° to 360° [indicated by the dark dashed circle in **Figure 3(b)**].

We projected the M-point of graphene BZ to the superimposed rotated 1st h-BN BZ, as the red point in **Figure 3(b-c)** indicated. The projected BZ point is located inside the h-BN BZ as shown in **Figure 3(c)**. Then we pinpoint the corresponding hopping starting point on the first valence band of *h*-BN. As indicated by the black dashed arc, along the graphene M-point projection trace in the *h*-BN BZ, the valence band energy decreases by approximately 0.8 eV from the M-point at 0° to a non-high-symmetry point on the Γ-K line at 30°. The energy drop is thus equivalent to the shift of the starting point projected on the *h*-BN valence band. We tracked the energy shift when the twist angle changes from 0° to 30° [gray line in **Figure 3(a)**], and it qualitatively matches the tendency of the experimental results.

For the 3rd transition appearing at 18°~25°, it may come from the band crossing of *h*-BN and graphene without symmetry protection at nearby angles. The accidental band crossings contribute to the vHS, which may result in the 3rd transition. Unfortunately, in our models, the lattice of h-BN is less than 2% compressed or tensioned to minimize the size of the commensurate moiré lattice to trade off the computational cost, while the lattice of graphene remains unchanged, and different amounts of lattice compression or tension will make the bandgap larger or smaller, respectively (**Figure S4**). At the same time, the band gaps are also under estimated due to the well know "band gap error" of the PBE functional. Both factors make the evaluation of the relative band structures as well as the band overlap between *h*-BN and graphene not quantitatively accurate, making tracking the features of the 3rd transition in the calculations challenging.

Notably, the difference of the transition energies enables identification of the twist angle by characteristic energies. In our experiment, the graphene we used is whole single crystalline[37], while the *h*-BN stacked on graphene is actually polycrystalline in micrometer scale[38], which results in some horizonal boundaries between *h*-BN/graphene heterostructures. **Figure 4(a)** shows a schematic of a typic horizontal boundary with different twist angles, which are estimated to be 27.6° and 17.5° by diffraction patterns [**Figure 4(b-c)**], but are difficult to distinguish in low-magnification

HAADF images. The typical EELS spectra are shown in **Figure 4(d)**. The graphene intralayer transition energy and h-BN/graphene interlayer transition energy show obvious redshifts of ~100 meV and ~280 meV from 17.5° to 27.6°, respectively, while the bandgap of h-BN remains unchanged (evaluated by the upturn before 6 eV). We can clearly distinguish the characteristic energies of the graphene intralayer transition and h-BN/graphene interlayer transition from the extracted energy distributions [**Figure 4(e-f)**], which show sharp contrasts at the boundary. For comparison, the mapping of the h-BN bandgap [**Figure 4(g)**] does not have any spatial features and remains uniform, indicating an intrinsic feature without influence from the moiré potential.

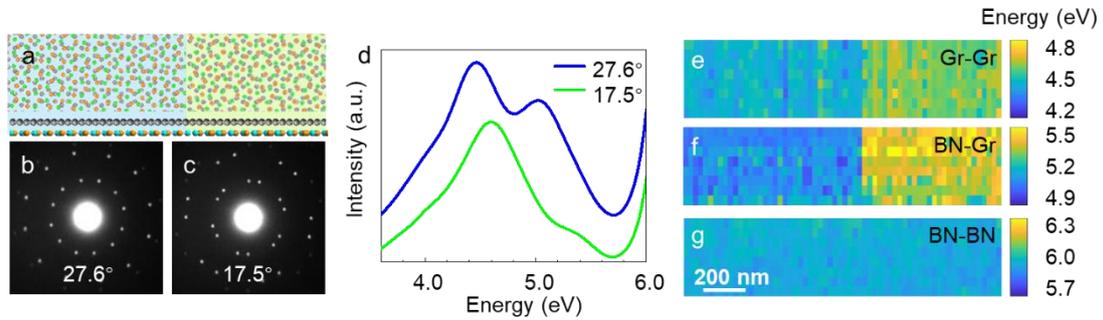

**Figure 4**. EELS mapping of the transition energies. (a) Schematic of a 27.6°/17.5° twist h-BN/Gr heterostructure boundary. (b-c) Diffraction patterns recorded on different sides of the boundary with a twist angle of (b) 27.6° (left) or (c) 17.5° (right). (d) Comparison between the integrated spectra on different sides of the boundary. (e-g) Energy mapping of the (e) Gr intralayer transition, (f) h-BN/Gr interlayer transition and (g) h-BN intralayer transition at the boundary.

In conclusion, we studied the twist angle-dependent intralayer and interlayer transition behavior in h-BN/graphene heterostructures. Owing to the moiré potential, the graphene M-point intralayer hopping energy decreases by ~0.25 eV when graphene contacts h-BN and when the twist angle between them increases from 0° to 30°. Meanwhile, the flat conduction band at the graphene M-point offers a new transition end state for the electrons in the h-BN valence band, making an interlayer transition with an energy even lower than the bandgap of h-BN possible. Because the transition starting point changes when graphene twists on h-BN while the end point is fixed at the graphene M-point, the transition energy varies from ~5.6 eV to ~5.1 eV when the twist angle increases from 0° to 30°. Our study reveals that in twisted 2D systems, even heterointerfaces between a wide bandgap insulator and a semimetal, such as *h*-BN and graphene, can contribute to new transition channels with different twist angles. Therefore, we should carefully address each 2D heterointerface to prevent unexpected influence on devices. Additionally, this provides new ways to design advanced 2D optoelectronic devices with artificial wavelengths.

The work is supported by the National Key R&D Program of China (2019YFA0708200 and 2019YFA0308500), the National Natural Science Foundation of China (52125307, 11974023, 52021006, T2188101, 52025023, 51991342, 92163206, 12104017), the "2011 Program" from the Peking-Tsinghua-IOP Collaborative Innovation Center of Quantum Matter, Youth Innovation Promotion Association, CAS. We acknowledge professor Yu Ye from Peking University for the helpful discussion, Electron Microscopy Laboratory of Peking University for the use of electron microscopes, and the High-performance Computing Platform of Peking University for providing computational resources.

# Supplemental Material

1. Methods

**CVD growth of monolayer graphene and *h*-BN.** For the growth of graphene, the industrial Cu foil (25 μm thick, 99.8%, Sichuan Oriental Stars Trading Co. Ltd.) was placed on a quartz substrate and then loaded into a CVD furnace. The Cu foil was then heated to 1030 °C under 500 sccm Ar and 10 sccm $H_2$. $CH_4$ (1–5 sccm) was introduced during the growth. After growth, the CVD system was cooled down to room temperature.

For the growth of *h*-BN, the precursor ammonia borane (97%; Aldrich) was filled into an $Al_2O_3$ crucible and placed at a distance of 1 m from a Cu foil substrate. First, the substrate was heated to the growth temperature (1,035 °C) under 500 sccm Ar and 50 sccm $H_2$ at atmospheric pressure. The CVD system was then switched to low pressure (about 200 Pa) under 5 sccm Ar and 45 sccm $H_2$, while the precursor was heated to 65 °C using a heating band. After growth, the CVD system was cooled down to room temperature.

**Transfer.** We used PMMA to help transferring mono layer Graphene and *h*-BN on holey carbon TEM grids. In the process, the PMMA was spin cast onto a graphene coated copper foil, and baked under 120°C for 3 min. With the protection of PMMA, the copper foil was etched away in 4% $(NH_4)_2S_2O_8$ solution for about 6 hours. Owing to surface tension, the graphene-PMMA film could float on the liquid surface and we used a TEM grid to pick it up. Then we baked the TEM grid under 120°C for 10 min to ensure the graphene stick tightly with the holy carbon. The PMMA on graphene was finally dissolved in acetone for 30 min. Another mono layer *h*-BN was transferred on the graphene TEM grid in the same way.

The *h*-BN/graphene heterostructure was obtained by transferring *h*-BN onto homemade monolayer graphene TEM grids using the PMMA-based transfer technique. The graphene TEM grids were prepared by transferring large-area monolayer single-crystal graphene on commercial holey-carbon-film TEM grids (Zhongjingkeyi GIG-2010-3C).

**STEM-EELS characterization.** The STEM-EELS data was acquired from a Nion U-HERMES200 STEM equipped with an aberration corrector and a monochromator operated at 60 kV. All the samples were baked at 160°C in vacuum for 8 hours before loaded into the microscope to remove the residual amorphous carbon on the sample and keep the column clean. The atomic resolution HAADF images and EELS were taken under 35 mrad and 20 mrad convergence semi-angles, respectively. The SAED were recorded under a parallel beam mode with a convergence semi-angle of 0.1 mrad. The EELS collection semi-angle was set to be 24.9 mrad. The energy resolution of the EELS could be decided with the full width at half maximum (FWHM) which is about 27.9 meV during the acquisition.

**Model constructions.** Commensurate $h$-BN/graphene heterostructures with twist angles in [0°, 30°] were generated by an ergodic searching algorithm, and the lattice constants in x-y plane are truncated by 20 Å. The lattice mismatch of $h$-BN is less than 2%, while the lattice of graphene keeps the same.

**DFT calculations.** Quantum mechanical calculations within the density functional theory was employed, as implemented in the Vienna ab initio simulation package (VASP) with the projector augmented wave (PAW) method. Generalized Gradient Approximations (GGA) in the form of PBE[1,2] was adopted for the exchange-correlation potential. The energy cutoff of the plane-wave basis set was 750 eV for GGA-level calculations. A vacuum layer thicker than 20 Å was used to avoid the image replica interactions between adjacent layers. Noninteracting band and DOS are calculated by simply superpose the $h$-BN onto graphene and the interlayer distance is set to 15 Å to avoid interlayer hoppings. Van der Waals corrections of the total energy was considered within the DFT-D3 method by Grimme et al[3,4]. A convergent Γ-centered 12 × 12 × 1 kpoint mesh was used for Brillouin zone sampling for the 21.8° configuration with ~6.53 Å lattice constants, and the similar kpoint density for other twist angles. Atomic coordinates are optimized with a conjugate gradient algorithm to ensure the energy difference and internal forces are less than $10^{-5}$ eV and 0.01 eV/Å, respectively.

## 2. Relationship between moiré wavelengths and twist angles

The relationship between the moiré wavelength and twist angle follows [5,6]:

$$L = \frac{p}{\sqrt{1+p^2-2p\cos\theta}} a \quad (1)$$

where $L$ is the moiré wavelength; $\theta$ is the twist angle; $a$ is the lattice constant of the substrate, equal to 2.505 Å for h-BN; and $p$ is the ratio between the lattice constants of the twist layers, equal to 0.982 for graphene and h-BN.

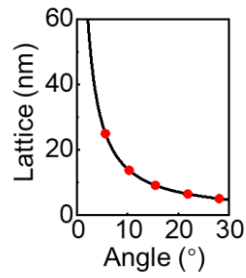

**Figure S1**. Relationship between the moiré wavelength and twist angle. The line is calculated from Eq. (1), and the twist angles and lattice constants of the red points are measured from **Figure 1(b)** and **Figure 1(c)**, respectively.

## 3. Diffraction patterns of h-BN/graphene heterojunctions

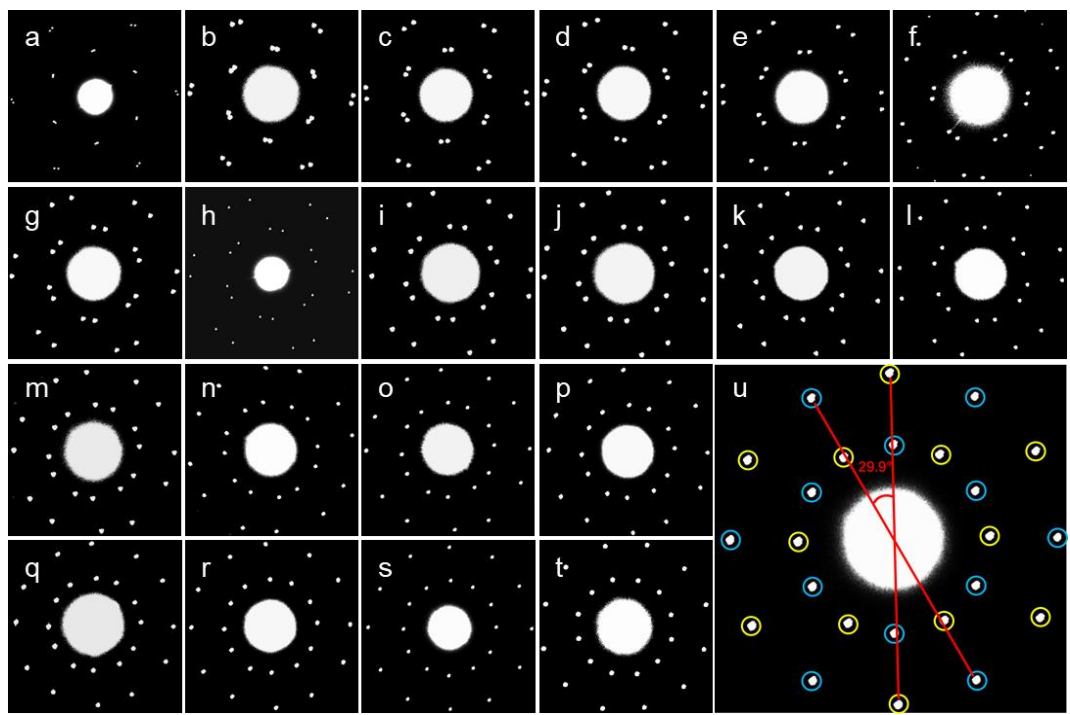

**Figure S2.** SAED patterns of h-BN/Gr with different twist angles [(a) to (u) corresponding to twist angles of 2.5°, 5.8°, 8.2°, 9.7°, 12.3°, 13.3°, 14.3°, 15.7°, 18.4°, 20.1°, 21.0°, 22.0°, 22.9°, 24.0°, 24.7°, 25.1°, 25.7°, 27.0°, 28.4°, 29.1° and 29.9°]. The slight length difference of the reciprocal lattices could be utilized to distinguish h-BN and graphene. For example, in (u), the spots in blue and yellow circles correspond to diffraction patterns of h-BN and graphene, respectively. The twist angle is ~29.9°.

## 4. Mechanism of the zero-momentum transfer measurement in EELS

The differential cross section in EELS satisfies[7]:

$$\frac{d^2\sigma}{d\Omega dE} \propto \frac{\text{Im}[-1/\varepsilon(q,E)]}{\theta^2+\theta_E^2} \quad (2)$$

where $\sigma$ is the cross section; $\Omega$ is the solid angle; $E$ is the energy loss in EELS; $\varepsilon$ is the dielectric function, which is a function of $E$ and momentum transfer $q$; $\theta$ is the scattering angle; and the characteristic angle $\theta_E = \frac{E}{\gamma m_0 v^2}$. For the 60 keV incident energy in our experiment, $\gamma m_0 v^2 = 113.7$ keV, in which $\gamma$ is the relativistic factor, $m_0$ is the electron rest mass, and $v$ is the electron velocity. We used $E \approx 5$ eV to represent the typical energy in our EELS measurement.

According to Ref.[8], the $\text{Im}[-1/\varepsilon(q,E)]$ in graphene or h-BN does not change in magnitude. Therefore, the main influence on the differential cross section comes from the $\frac{1}{\theta^2+\theta_E^2}$ factor. Thus, the momentum transfer distribution in our EELS measurement is proportional to:

$$\int_{-\alpha}^{\alpha} f(a) \cdot \frac{1}{\theta^2+\theta_E^2} w(a-\theta) da \quad (3)$$

in which $a$ is the e-beam incident angle, $f(a)$ is the incident angle distribution, and the term

$$w(a-\theta) = \begin{cases} 1, & -\beta \leq a-\theta \leq \beta \\ 0, & a-\theta < -\beta \text{ or } a-\theta > \beta \end{cases} \quad (4)$$

determines whether the electrons are scattered within the range of the collective semi-angle β.

In our EELS measurement, the convergence semi-angle α=20 mrad, and the collective semi-angle β=24.9 mrad. With the assumption that the incident beam has a uniform distribution within the convergence semi-angle, $f(a) = 1$, we have:

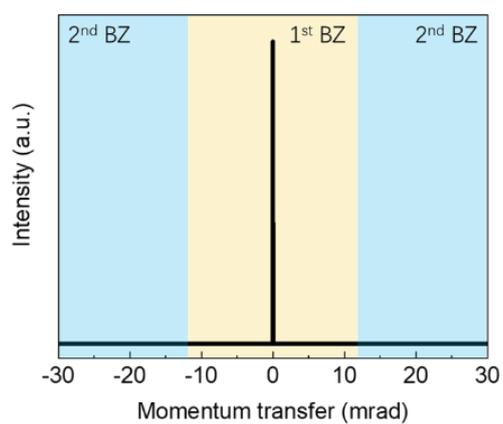

**Figure S3.**   Distribution of the momentum transfer in our EELS measurement.

As we can see in **Figure S3**, the momentum transfer distribution is almost divergent at 0, which means that our EELS experiment mainly records the direct transition without momentum transfer.

## 5. The band shift caused by the lattice distortion in calculation

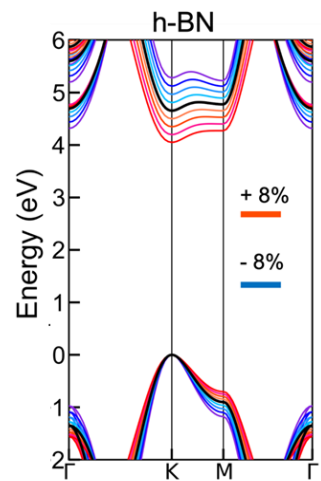

**Figure S4.** Band structure of h-BN with up to ±8% lattice mismatch; the step size is 2%.